\documentclass[a4paper]{jpconf}
\usepackage{graphicx}
\begin{document}

{\ }
\vspace{10cm}
{\ }

This is an unedited preprint. The original article is published under licence in {\sl Journal of Physics: Conference Series} by IOP Publishing Ltd and is available at \\ \ \\

\centerline{http://iopscience.iop.org/1742-6596/217/1/012125}

\ 

\centerline{http://www.doi.org/10.1088/1742-6596/217/1/012125}

\newpage

\title{Pressure-induced changes of the vibrational modes of spin-crossover complexes studied by nuclear
       resonance scattering of synchrotron radiation}

\author{A X Trautwein$^1$, H Paulsen$^1$, H Winkler$^1$, H Giefers$^2$, G Wortmann$^2$, H Toftlund$^3$, J A Wolny$^4$, A I Chumakov$^5$ and O Leupold$^5$}

\address{$^1$ Institut f\"{u}r Physik, Universit\"{a}t zu L\"{u}beck, D-23538 L\"{u}beck, Germany}
\address{$^2$ Department Physik, Universit\"{a}t Paderborn, D-33095 Paderborn, Germany}
\address{$^3$ Department of Chemistry, University of Southern Denmark, Odense, DK-5230, Denmark}
\address{$^4$ Fachbereich Physik, Technische Universit\"{a}t Kaiserslautern, D-67663 Kaiserslautern, Germany}
\address{$^5$ European Synchrotron Radiation Facility, BP 220, F-38042 Grenoble, France}

\ead{trautwein@physik.uni-luebeck.de}

\begin{abstract}
Nuclear inelastic scattering (NIS) spectra were recorded for the spin-crossover complexes STP and ETP (STP = [Fe(1,1,1-tris\{[N-(2-pyridylmethyl)-N-methylamino]methyl\}\-ethane)](ClO$_4$)$_2$ and ETP = [Fe(1,1,1-tris\{[N-(2-pyridylmethyl)-N-methylamino]methyl\}\-butane)](ClO$_4$)$_2$) at 30 K and at room temperature and also at ambient pressure and  applied pressure (up to 2.6 GPa).
Spin transition from the high-spin (HS) to the low-spin (LS) state was observed by lowering temperature and also by applying pressure at room temperature and has been assigned to the hardening of iron-bond stretching modes due to the smaller volume in the LS isomer.
\end{abstract}

\section{Introduction}
The iron(II) complexes STP and ETP (STP = [Fe(1,1,1-tris\{[N-(2-pyridylmethyl)-N-methylamino]methyl\}ethane)](ClO$_4$)$_2$ and ETP = [Fe(1,1,1-tris\{[N-(2-pyridylmethyl)-N-methylamino]methyl\}butane)](ClO$_4$)$_2$) belong to the family of thermally driven spin-crossover complexes, which exhibit a transition from the low-spin (LS, $S=0$) to the high-spin (HS, $S=2$) state by increasing the temperature.
These complexes are promising materials for optical information storage and display devices \cite{Guetlich1994}.
The first coordination sphere of the iron in STP and ETP is a distorted [FeN$_6$] octahedron with $C_3$ symmetry (Fig. \ref{fig1}).

The spin transition occurs in both complexes gradually at about 180 K.
Its driving force is the difference of the vibrational entropy between both spin states, which arises from softening of some vibrational modes when going from the LS to the HS state.
NIS measurements on STP have demonstrated that mainly the iron-ligand bond stretching modes are affected this way \cite{Paulsen2001,Paulsen2001a}.
Similar findings were obtained for other spin-crossover complexes \cite{Paulsen1999,Boettger2006,Ronayne2006}.
Spin transition is also achieved by increasing pressure; thus the LS becomes energetically favoured over the HS state due to its smaller volume.

\section{Nuclear inelastic scattering (NIS)}

\begin{figure}[htb]
\begin{minipage}{17.5pc}
\centerline{\includegraphics[width=11.5pc]{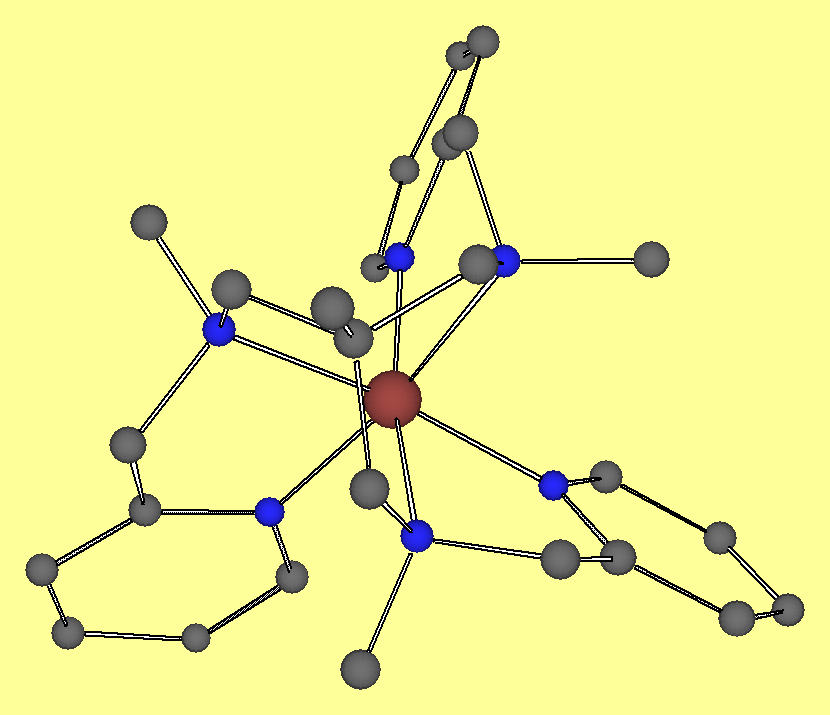}}
\caption{\label{fig1}Molecular structure of STP.}
\vspace{2pc}
\includegraphics[width=17.5pc]{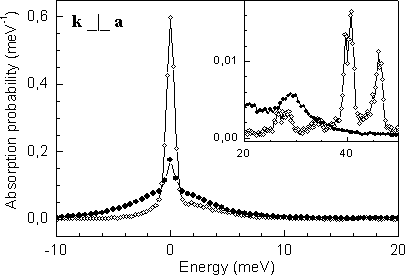}
\caption{\label{fig5a}NIS spectra measured at $T=30$ K ($\diamond$) and at room temperature ($\bullet$) for
         $\vec{k}\perp\vec{a}$.
         The inset shows the energy range of the iron-ligand bond stretching modes.
         The solid lines are guides to the eyes.
         Taken from \cite{Paulsen2001}.}
\end{minipage} 
\hfill
\begin{minipage}{17.5pc}
\includegraphics[width=17.5pc]{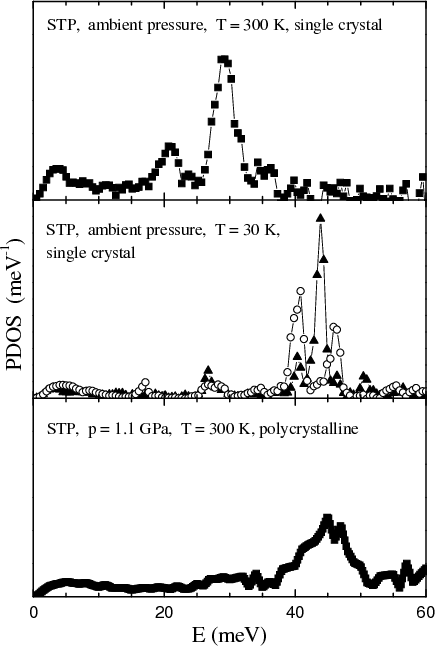}
\caption{\label{fig10}PDOS of STP for temperatures and pressures as indicated.
         The PDOS at 30 K was measured at two different orientations: $\vec{k}\perp\vec{a}$ (circles) and
         $\vec{k}\parallel\vec{a}$ (triangles).}
\end{minipage} 
\end{figure}

Contributions in the spectrum with negative energy $E$ account for annihilation and positive energy for creation of vibrations in the molecular complex (note that zero energy in Fig. \ref{fig5a} corresponds to the elastic part of the spectrum, while negative and positive energies cover the inelastic part).
At low temperatures only low-energy phonon states are occupied (2.5 meV corresponds to approximately 30 K), therefore annihilation of phonons is scarce in this region.
Creation of phonons, however, is possible at all vibrational energies.

The NIS spectra recorded for STP at 30 K and at room temperature (Fig. \ref{fig5a}) reflect in their inelastic part different iron-ligand bonding: at 30 K the Lamb-M\"{o}ssbauer factor $f_{\rm LM}$ is close to one ($f_{\rm LM}\approx 0.88$), thus dominating the elastic peak for the LS isomer, while at room temperature $f_{\rm LM}$ decreases to 0.16 yielding only a small elastic peak.
The inelastic part of the two NIS spectra (inset of Fig. \ref{fig5a}), at the same time, exhibits the drastic hardening of vibrational modes when decreasing temperature.

\section{Angular-resolved NIS measurements}

The high energy resolution (0.65 meV) available at the Nuclear Resonance Beamline ID18 of the European Synchrotron Radiation Facility (ESRF) in Grenoble, France, allowed us to resolve individual molecular vibrations which were unambiguously identified by DFT calculations \cite{Paulsen2001}.

\begin{figure}[htb]
\begin{minipage}{17.5pc}
\includegraphics[angle=-0.6,width=12pc]{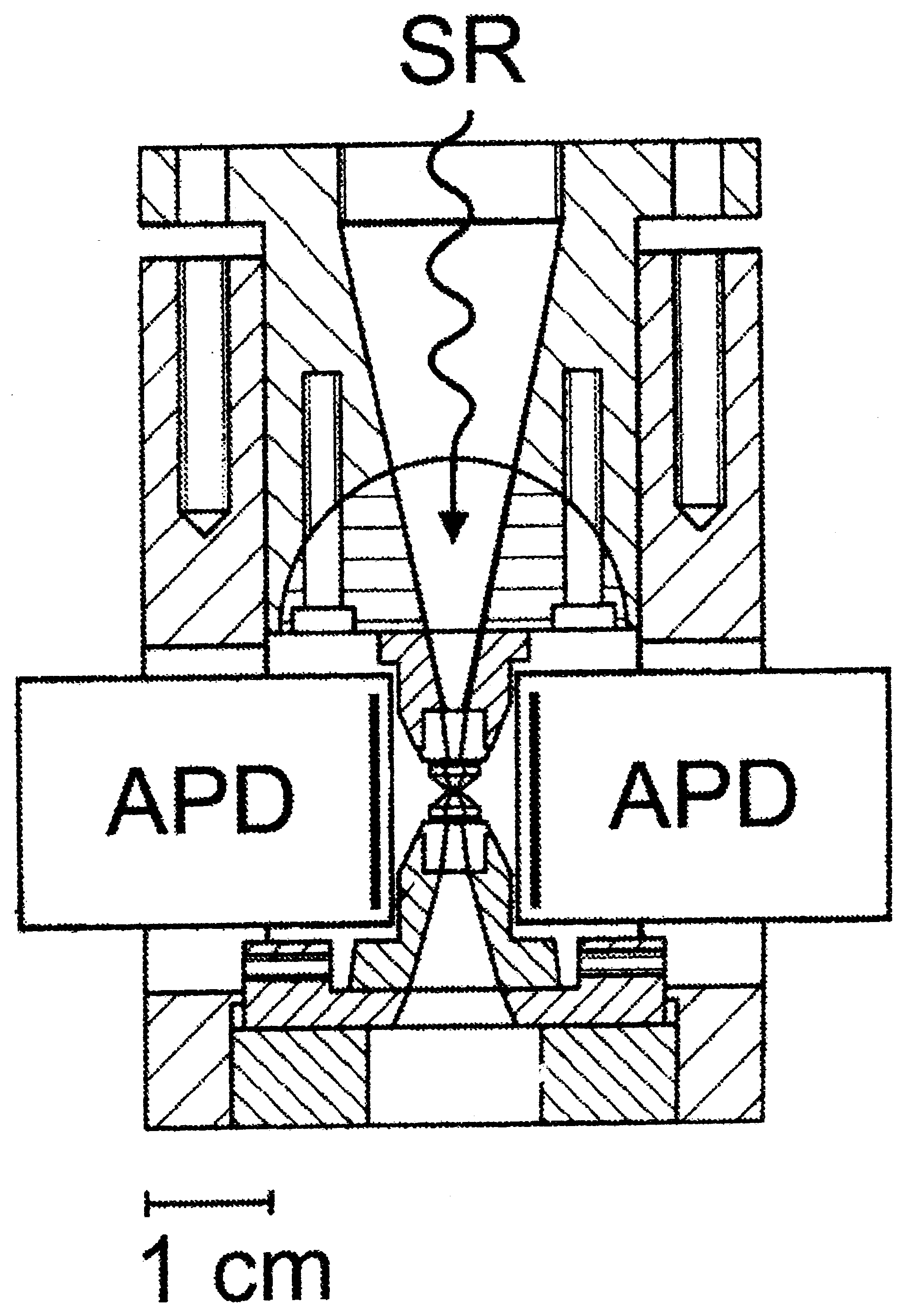}
\caption{\label{fig7}Diamond-anvil cell for high-pressure NFS and NIS studies.}
\end{minipage}
\hfill
\begin{minipage}{17.5pc}
\includegraphics[width=17.5pc]{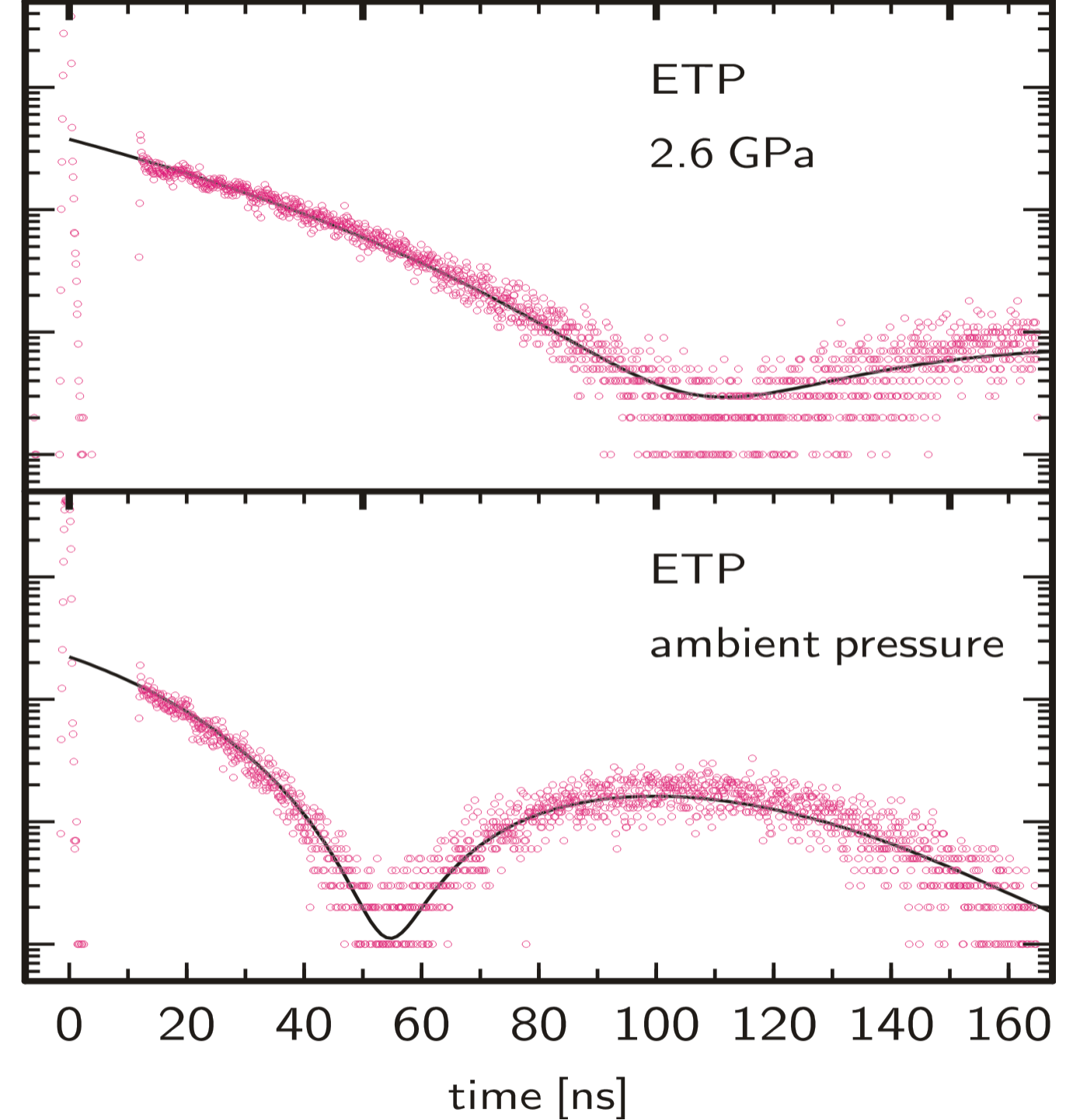}
\caption{\label{fig20}NFS spectra of ETP recorded at room temperature with ambient pressure and applied pressure as indicated.}
\end{minipage} 
\end{figure}

In case of molecular crystals, and if the interactions between {\it inter}- and {\it intra}-molecular modes can be neglected, the molecular part of the anisotropic PDOS (partial density of states) can be formulated as
\begin{displaymath}
  {\rm PDOS}_{\rm mol} = \sum_j \delta(E-\hbar \omega_j)(\vec{k}\cdot\vec{u}_j)^2
\end{displaymath}
with $\omega_j$ representing the vibrational frequencies of the molecule and $\vec{u}_j$ the projection of the $j$th eigenvector into the three-dimensional subspace of the iron coordinates, and $\vec{k}$ is the normalized wave vector of the photon.
From this expression it is obvious that only those vibrational modes will be NIS visible that are connected with a mean square displacement of the iron nucleus along the wave vector $\vec{k}$.

In Fig. \ref{fig10} the angular-resolved vibrational modes, recorded at 30 K for STP, are clearly visible.
For the interpretation of the experimentally gained PDOS, DFT calculations were performed.
This way the two predominant peaks at 40 and 46 meV for $\vec{k}\perp\vec{a}$ as well as the predominant peak at 44 meV and the small peak at 41 meV for $\vec{k}\parallel \vec{a}$ could be assigned to iron-ligand bond-stretching modes.
         
Considering an ideal [FeN$_6$] octahedron there exist six Fe-N bond-stretching modes which transform according to the $A_{1g}$, $E_g$ and $T_{1u}$ irreducible representations of the point group $O_h$.
The {\it gerade} modes $A_{1g}$ and $E_g$ are not connected with a displacement of the iron nucleus, and thus only the three $T_{1u}$ bond-stretching modes can be detected by NIS.
Actually, the first coordination sphere of the iron in the STP complex has to be regarded as a distorted [FeN$_6$] octahedron with $C_3$ symmetry.
Because of the lower symmetry the vibrational $T_{1u}$ term is split into an $A$ and a doubly degenerate $E$ term.
In this case all six bond-stretching modes are visible by NIS: two pairs of doubly degenerate $E$ modes are connected with a displacement of the iron nucleus within the equatorial plane and are visible if $\vec{k}$ lies within this plane that is perpendicular to the threefold symmetry axis [Fig. \ref{fig10} (open circles)].
The remaining two $A$ modes are connected with a displacement of the iron nucleus along the molecular symmetry axis [Fig. \ref{fig10} (filled triangles].
The smaller peak at 41 meV corresponds to the nearly fully symmetric breathing mode where the iron nucleus is only slightly participating in the vibration.

\section{Pressure-induced spin transition}

\begin{figure}[htb]
\begin{minipage}{17.5pc}
\centerline{\includegraphics[width=12pc]{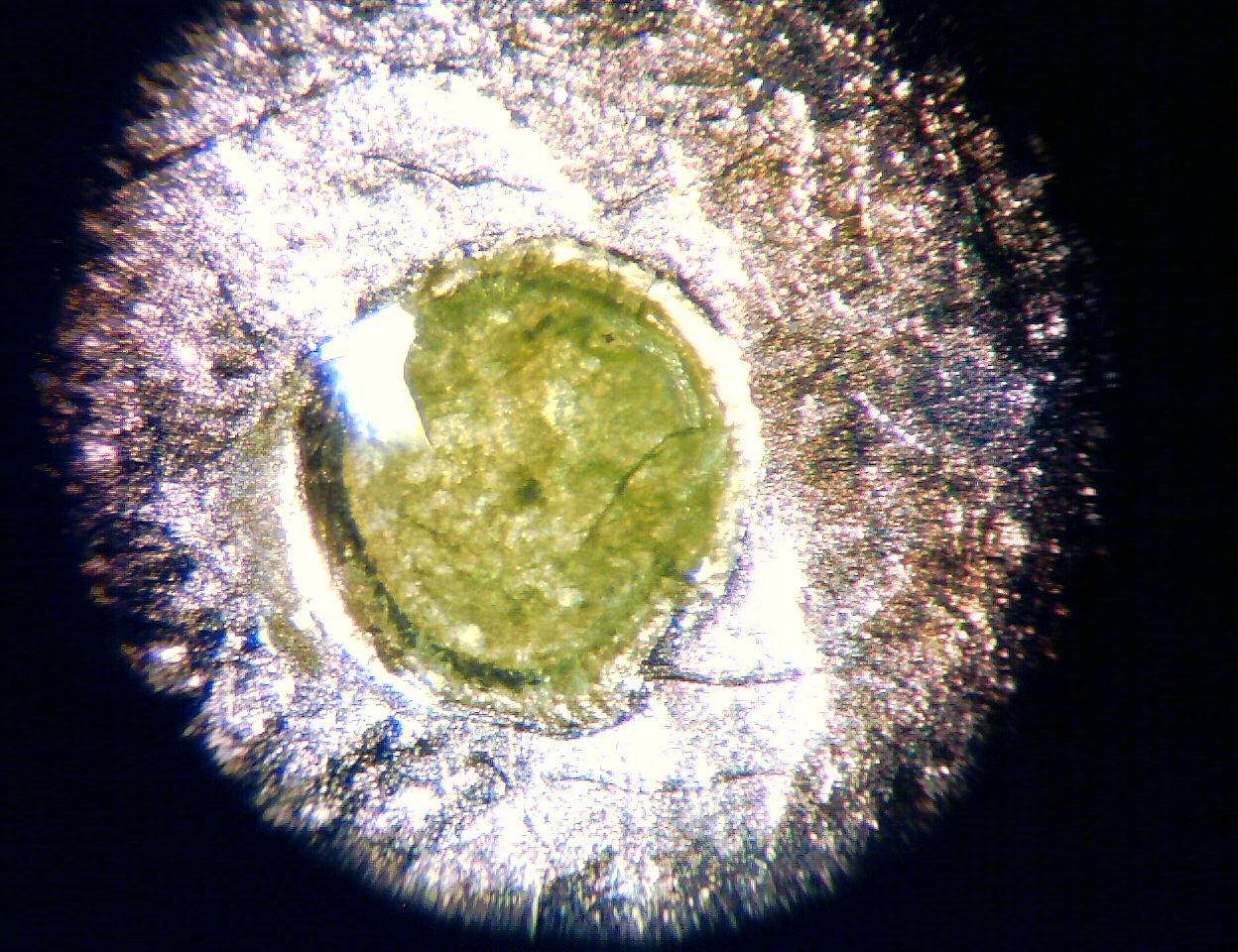}}
\vspace{0.4pc}
\end{minipage} 
\hfill
\begin{minipage}{17.5pc}
\centerline{\includegraphics[width=12pc]{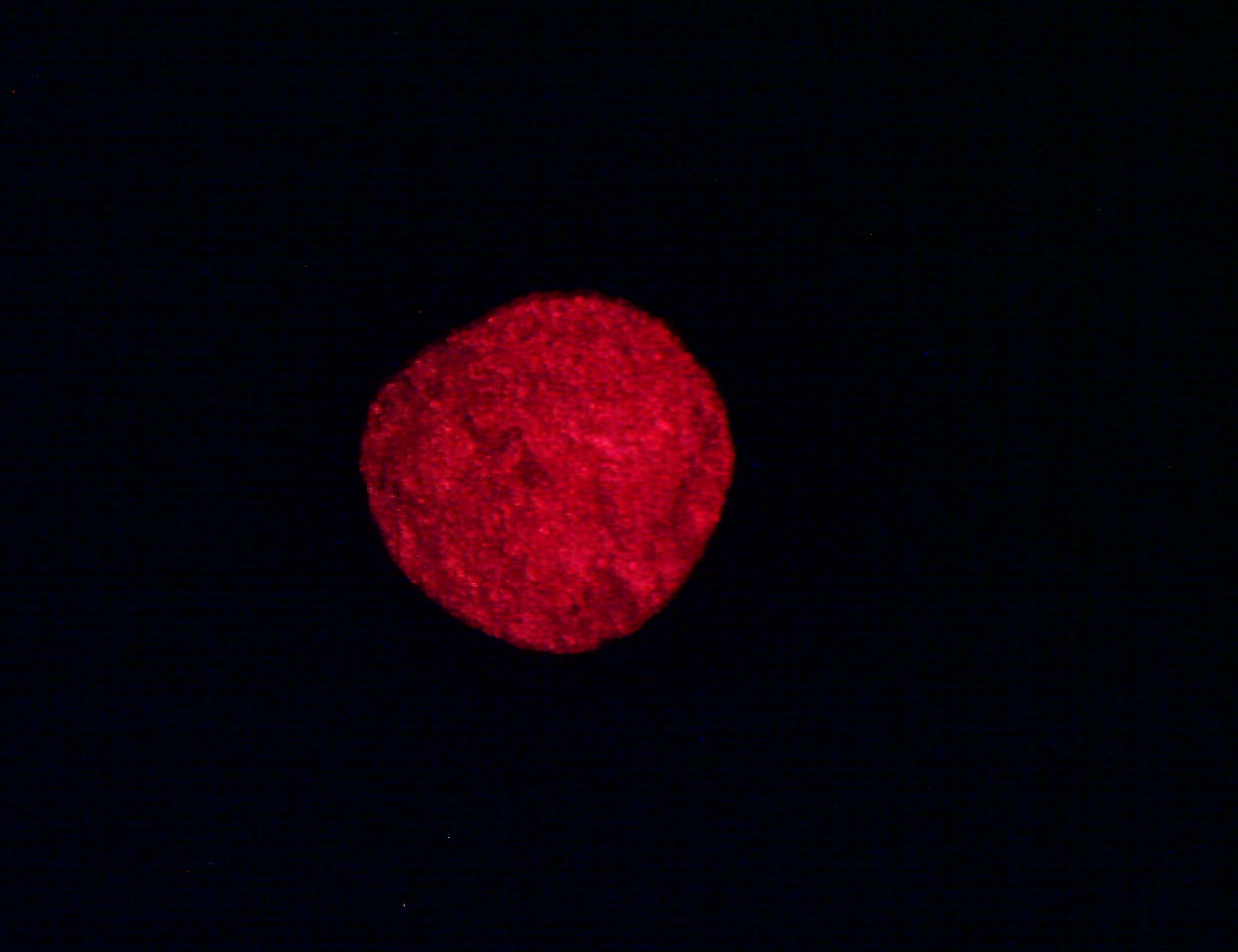}}
\vspace{0.4pc}
\end{minipage} 
\caption{\label{fig5b} Colour change of the STP sample at ambient pressure (left picture) and at applied pressure of 1.1 GPa (right picture).}
\end{figure}

Spin transition is also achieved by pressure, because with increasing pressure the LS state is energetically favoured due to its smaller volume.
The NFS and NIS spectra were recorded at ambient pressure and also at pressures up to 2.6 GPa in a diamond-anvil cell
(DAC, Fig. \ref{fig7}).
The NFS spectra of STP and ETP at ambient pressure exhibit the first quantum-beat minimum at about 55 ns (corresponding to $\Delta E_Q=0.8$ mm/s) indicative of the HS state.
When applying a pressure above 0.9 GPa, both STP and ETP samples undergo HS-LS transitions, with the quadrupole splitting of the LS isomer changing to a smaller value compared to that of the HS state (Fig. \ref{fig20}).
The spin transition was also monitored optically in the DAC by a characteristic change of the sample colour from green to red (Fig. \ref{fig5b}).
\vspace{0.9pc}
The PDOS of Fe in STP, extracted from the NIS spectra measured at 1.1 GPa and 300 K, is shown in Fig. \ref{fig10} together with the PDOS obtained from a single crystal of STP.
The comparison with the PDOS at ambient pressure and 300 K (HS state) and at 30 K (LS state) strikingly demonstrates that STP at 1.1 GPa and 300 K has transformed into the LS phase with comparable frequencies of the high-energy stretching modes (note that the PDOS of STP at 30 K in Fig. \ref{fig10} is shown for two different directions (filled triangles, open circles) while for the polycrystalline sample all directions are averaged).

\section*{Acknowledgments}
The authors acknowledge the support by R. R\"{u}ffer and by the relevant ESRF services and the financial support by the Deutsche Forschungsgemeinschaft (DFG) and by the German Federal Ministery for Education and Research (BMBF).

\end{document}